\documentclass[twocolumn]{aastex63}
\usepackage{amssymb, amsmath}
\usepackage{graphicx}
\usepackage{natbib}
\usepackage{latexsym}
\usepackage{mathrsfs}
\usepackage{mathtools}
\usepackage{multirow}
\usepackage{xcolor}
\usepackage{color, colortbl}
\usepackage{booktabs,makecell}

\newcommand{\tsup}{\textsuperscript}
\newcommand{\tsub}{\textsubscript}

\received{June 15, 2020}
\revised{August 11, 2020}
\accepted{August 12, 2020}
\submitjournal{AJ}

\begin{document}

\title{Optical Transmission Spectroscopy of the Terrestrial Exoplanet LHS 3844b \\from 13 Ground-Based Transit Observations}

\correspondingauthor{Hannah Diamond-Lowe}
\email{hdiamondlowe@cfa.harvard.edu}

\author[0000-0001-8274-6639]{Hannah Diamond-Lowe}
\affil{Center for Astrophysics $\vert$ Harvard \& Smithsonian, 60 Garden St., Cambridge, MA 02138, USA}

\author[0000-0002-9003-484X]{David Charbonneau}
\affiliation{Center for Astrophysics $\vert$ Harvard \& Smithsonian, 60 Garden St., Cambridge, MA 02138, USA}

\author{Matej Malik}
\affiliation{Department of Astronomy, University of Maryland, College Park, MD 20742, USA}

\author[0000-0002-1337-9051]{Eliza M.-R. Kempton}
\affiliation{Department of Astronomy, University of Maryland, College Park, MD 20742, USA}

\author{Yuri Beletsky}
\affiliation{Las Campanas Observatory, Carnegie Institution of Washington, Colina el Pino, 601 Casilla, La Serena, Chile}
 
\begin{abstract}

Atmospheric studies of spectroscopically accessible terrestrial exoplanets lay the groundwork for comparative planetology between these worlds and the Solar System terrestrial planets. LHS 3844b is a highly-irradiated terrestrial exoplanet ($R=1.303\pm0.022R_{\oplus}$) orbiting a mid-M dwarf 15 parsecs away. Work based on near-infrared Spitzer phase curves ruled out atmospheres with surface pressures $\geq10$ bars on this planet. We present 13 transit observations of LHS 3844b taken with the Magellan Clay telescope and the LDSS3C multi-object spectrograph covering 620-1020 nm. We analyze each of the 13 data sets individually using a Gaussian process regression, and present both white and spectroscopic light curves. In the combined white light curve we achieve an RMS precision of 65 ppm when binning to 10-minutes. The mean white light curve value of $(R_p/R_s)^2$ is $0.4170\pm0.0046$\%. To construct the transmission spectrum, we split the white light curves into 20 spectrophotometric bands, each spanning 20 nm, and compute the mean values of $(R_p/R_s)^2$ in each band. We compare the transmission spectrum to two sets of atmospheric models. We disfavor a clear, solar composition atmosphere ($\mu=2.34$) with surface pressures $\geq$0.1 bar to 5.2$\sigma$ confidence. We disfavor a clear, H$_2$O steam atmosphere ($\mu=18$) with surface pressures $\geq$0.1 bar to low confidence (2.9$\sigma$). Our observed transmission spectrum favors a flat line. For solar composition atmospheres with surface pressures $\geq$1 bar we rule out clouds with cloud-top pressures of 0.1 bar (5.3$\sigma$), but we cannot address high-altitude clouds at lower pressures. Our results add further evidence that LHS 3844b is devoid of an atmosphere.\\
\end{abstract}

\section{Introduction} \label{sec:intro}

Like the terrestrial planets of the Solar System, terrestrial exoplanets have radii $R<1.6R_{\oplus}$ and bulk densities that imply iron cores surrounded by rocky mantles. As yet we do not know what the atmospheres around these worlds look like, or if they bare any similarity to the high mean molecular weight secondary atmospheres that surround Venus, Earth, and Mars, or the tenuous envelope around Mercury. Terrestrial exoplanets are distinct from another class of small planets, usually referred to as mini-Neptunes \citep{Owen&Wu2013,Lopez2013,Rogers2015,Dressing2015a,Fulton2017,VanEylen2018}. These worlds are consistent with iron-rock interiors surrounded by thick envelopes of hydrogen- and helium-dominated gas, and are unlike any planets we see in the Solar System.

Current instrumentation allows for atmospheric follow-up of mini-Neptunes \citep[e.g.,][]{Kreidberg2014,Benneke2019}, but the small signals produced by terrestrial exoplanets make in-depth studies of most of their atmospheres out of reach for our telescopes. The most spectroscopically accessible terrestrial exoplanets orbit nearby ($<15$ pc), small ($<0.3R_{\odot}$) stars. Ground-based surveys, like MEarth \citep{Nutzman2008,Irwin2015} and TRAPPIST \citep{Gillon2013}, and the space-based Transiting Exoplanet Survey Satellite \cite[TESS;][]{Ricker2015} have compiled a small sample of terrestrial exoplanets that meet these requirements. 

One such terrestrial exoplanet is the highly irradiated world LHS 3844b \citep{Vanderspek2019}, first identified with TESS. As of this writing there is no published mass for LHS 3844b, but its radius of $R=1.303\pm0.022R_{\oplus}$ places it squarely in the radius regime of terrestrial planets. LHS 3844b is the third in a series of four terrestrial exoplanets whose atmospheres we address with ground-based transmission spectroscopy. For the terrestrial exoplanet GJ 1132b \citep{Berta-Thompson2015} we disfavor a clear, low mean molecular weight atmosphere \citep{Diamond-Lowe2018}, while the atmosphere of the habitable-zone terrestrial planet LHS 1140b \citep{Dittmann2017a,Ment2019} is below the detection limits of our instruments \citep{Diamond-Lowe2020a}. A data set on the nearby terrestrial planet orbiting LTT 1445A \citep{Winters2019} is forthcoming. Clear, low mean molecular weight atmospheres are also ruled out for five of the seven TRAPPIST-1 planets \citep{Gillon2016,Gillon2017a,deWit2016,deWit2018}.

It is an outstanding question whether or not terrestrial worlds orbiting M dwarfs can retain atmospheres at all. Unlike their solar-type counterparts, M dwarfs spend more time in the pre-main sequence phase \citep{Baraffe2015} during which they exhibit enhanced magnetic activity and emit high levels of damaging ultra-violet and X-ray radiation. This high energy radiation can drive atmospheric mass loss, as well as alter the photochemistry of any remaining atmosphere \citep{France2013}. LHS 3844b orbits so close to its host star, with an orbital period of 11 hours \citep[$T_{eq}$=805 K, $S$=70$S_{\oplus}$;][]{Vanderspek2019}, that any atmosphere around this world has likely been driven away via photodissociation and hydrodynamic escape \citep{Tian2014,Luger2015,Rugheimer2015}.

Using 100 hours of almost continuous observations with the Spitzer Space Telescope, \citet{Kreidberg2019} observed nine orbits of LHS 3844b to determine whether or not this world has a thick atmosphere. Short-period terrestrial planets like LHS 3844b are tidally locked \citep{Kasting1993}, so energy advection from the day-side to the night-side can only occur through atmospheric transport, with thicker atmospheres more efficient at doing so \citep{Showman2013,Wordsworth2015,Koll2019}. Phase curve information can reveal evidence of energy advection if there is an offset in the peak of the phase curve from the substellar point, and if the peak-to-trough variation is smaller than predicted for a bare rock. \citet{Kreidberg2019} found a day-side brightness temperature of 1040 $\pm$ 40 K and a night-side temperature consistent with 0 K for LHS 3844b, which rules out atmospheres with surface pressures $\geq10$ bar. Based on theoretical calculations the authors argue that more tenuous atmospheres, those with surface pressures $<1$ bar, are not stable to the high energy radiation from LHS 3844 over the planet's lifetime. 

\citet{Kreidberg2019} used Channel 2 of Spitzer's IRAC camera, which has a broad photometric band of 4-5$\mu$m, to gather phase curve and emission data of LHS 3844b. In this work we use the Magellan II (Clay) telescope and the LDSS3C multi-object spectrograph to gather optical spectra from 620-1020 nm of LHS 3844 before, during, and after the planet transit. With this data we employ the technique of transmission spectroscopy to address the atmosphere of LHS 3844b.

This paper is laid out as follows: In Section~\ref{sec:obs} we detail our observing program. We briefly describe our extraction pipeline and analysis in Section~\ref{sec:extractanalysis} \citep[a more detailed description has already been published in][]{Diamond-Lowe2020a}. We present our results along with a discussion in Section~\ref{sec:resultsdiscussion}. Our conclusions can be found in Section~\ref{sec:conclusion}.

\section{Observations} \label{sec:obs}

LHS 3844b orbits rapidly about its mid-M dwarf host, with an orbital period of 0.4629279 $\pm$ 0.0000006 days (11.11 hours) and a transit duration of 0.02172 $\pm$ 0.00019 days (31.3 minutes) \citep{Vanderspek2019}. Transits of LHS 3844b occur frequently and easily fit within an observing night. However, the signal-to-noise is proportional to the square root of the number of in-transit photons that are detected, so the short transit duration means that we must stack many transits together in order to build up the signal. 

Between June and October of 2019, the Center for Astrophysics $\vert$ Harvard \& Smithsonian awarded us 18 opportunities to observe transits of LHS 3844b with the Magellan II (Clay) telescope and the Low Dispersion Survey Spectrograph (\href{http://www.lco.cl/?epkb_post_type_1=ldss-3-user-manual}{LDSS3C}) at the Las Campanas Observatory in Chile (PI Diamond-Lowe). Each observation opportunity comprised 2.5 hours to observe the LHS 3844b transit, along with baseline on either side with which to remove systematic noise and measure the transit depth. Of the 18 opportunities, 13 resulted in data sets that we use in our analysis (Table~\ref{tab:obs}).

\begin{deluxetable*}{cccccccccl}
\centering
\caption{Observations with Magellan II (Clay) and the LDSS3C Multi-Object Spectrograph\label{tab:obs}}
\tablewidth{0pt}
\tablehead{
\colhead{Data Set} & \colhead{Transit} & \colhead{Night} & \colhead{Time} & \colhead{Exposure Time} & \colhead{Duty Cycle} & \colhead{Number of} & \colhead{Minimum} & \colhead{Median Seeing} & \colhead{Observing}\\
\colhead{Number}   & \colhead{Number\tsup{a}} & \colhead{ (UTC; 2019) }    & \colhead{(UTC)} & \colhead{(s)}           & \colhead{(\%) }& \colhead{Exposures} & \colhead{Airmass}        & \colhead{(arcsec)} & \colhead{Conditions\tsup{b}}
}
\startdata
* & 698 & 06-14 & -------------------------- & --- & --- & --- & ------ & --- & ---\\
$\dagger$ & 700 & 06-15 & -------------------------- & --- & --- & --- & ------ & --- \\
1   & 711 & 06-19 & 07:49:47 -- 10:11:36 & 30 & 63.8 & 177 & 1.308 & 1.1 & TCC \\
2   & 713 & 06-21 & 05:58:56 -- 08:09:05 & 30 & 63.8 & 167 & 1.340 & 1.0 & tCC \\
3   & 715 & 06-22 & 04:13:42 -- 06:21:48 & 30 & 63.8 & 165 & 1.491 & 1.5 & tCC \\
4   & 741 & 07-04 & 05:27:48 -- 07:04:30 & 30 & 63.8 & 123 & 1.352 & 0.9 & Cl, LW\\
5   & 765 & 07-15 & 07:48:43 -- 09:47:45 & 27 & 61.4 & 163 & 1.308 & 0.8 & Cl, LW\\
6   & 767 & 07-16 & 05:47:09 -- 07:58:02 & 27 & 61.4 & 179 & 1.308 & 0.6 & Cl, LW\\
7   & 769 & 07-17 & 04:12:47 -- 06:17:07 & 27 & 61.4 & 170 & 1.349 & 1.0 & Cl, LW\\
8   & 808 & 08-04 & 05:20:53 -- 07:34:27 & 27 & 61.4 & 184 & 1.308 & 0.6 & tCC\\
9   & 810 & 08-05 & 03:27:11 -- 05:48:52 & 27 & 61.4 & 195 & 1.317 & 0.6 & Cl, LW\\
10  & 812 & 08-06 & 01:39:09 -- 03:59:24 & 27 & 61.4 & 193 & 1.425 & 0.8 & Cl, LW\\
11  & 821 & 08-10 & 05:44:13 -- 07:54:41 & 27 & 61.4 & 179 & 1.308 & 1.1 & Cl, HW\\
12  & 825 & 08-12 & 02:25:08 -- 04:24:47 & 27 & 61.4 & 164 & 1.359 & 2.0 & Cl, LW\\
$\dagger$ & 834 & 08-16 & -------------------------- & --- & --- & --- & ------ & --- & ---  \\
13  & 838 & 08-18 & 02:52:19 -- 04:49:56 & 27 & 61.4 & 161 & 1.321 & 1.0 & tCC, lW\\
$\dagger$ & 879 & 09-06 & -------------------------- & --- & --- & --- & ------ & --- & ---\\
$\dagger$ & 881 & 09-07 & -------------------------- & --- & --- & --- & ------ & --- & ---\\
\enddata
\tablecomments{\\
* Observation lost due to instrumental problems\\
$\dagger$ Observations not taken due to bad weather \\
\tsup{a} Transit number is counted from the transit ephemeris $T_0 = 2458325.72559$ \citep{Kreidberg2019} \\
\tsup{b} Key: Cl = Clear sky; tCC = thin cirrus clouds; TCC = thick cirrus clouds; LW = low wind; HW = high wind
}
\end{deluxetable*}

LDSS3C has a single CCD detector with 15$\mu$ pixels arranged in a 2048 $\times$ 4096 configuration. Two amplifiers are used to read out the detector and convert incoming photons to electrons. The detector has a full well of 200,000 e$^-$, but the 16-bit analog-to-digital converter (ADC) saturates at 65,536 analog-to-digital units (ADUs). We ensure that no pixels used in our analysis exceed this saturation limit. We use the \texttt{Fast} readout mode with the \texttt{Low} gain setting and $2\times2$ binning to optimize the duty cycle.

LDSS3C was upgraded in September of 2014 to enhance sensitivity in the red optical part of the spectrum \citep{Stevenson2016a}, where M dwarfs emit the bulk of their photons. This is the primary reason we chose LDSS3C for these observations. We use the VPH-Red grism to observe over the nominal wavelength range of 620-1020 nm. Our observing program on LDSS3C is similar to ones we employed for GJ 1132b and LHS 1140b, two terrestrial exoplanets also transiting nearby mid-M dwarfs \citep{Diamond-Lowe2018,Diamond-Lowe2020a}. We achieve a duty cycle of 63.8\% or 61.4\% for all 13 of our observations of LHS 3844b.

Observing with a ground-based spectrograph means that telluric features are imprinted on stellar spectra before they reach the detector. Variations in precipitable water vapor translate into variations in the spectra of LHS 3844 large enough to wash out the signal we are looking for from LHS 3844b's atmosphere; telluric signatures can induce variations in the raw light curve of LHS 3844 by as much as 80\% on a night with variable weather conditions, while the largest features we might observe in the atmosphere of LHS 3844b produce variations of 0.04\%. We are able to compensate for the telluric variations with LDSS3C, a multi-object spectrograph with which we can simultaneously observe spectra of the target star, LHS 3844, and of comparison stars. The comparison stars are used to calibrate the LHS 3844 light curve, and they should be at least as bright as the target star so as to not be the photon-limiting factors, though not so bright as to bring down the duty cycle. The comparison stars must be selected before the observations so that an LDSS3C mask can be cut with slits corresponding to LHS 3844 and the comparison stars. We used the same calibration and science masks for all observations used in the analysis, and were able to achieve pointings consistent to within 10 pixels in the dispersion and cross-dispersion directions from night to night.

\begin{figure}[!ht]
\includegraphics[width=0.47\textwidth]{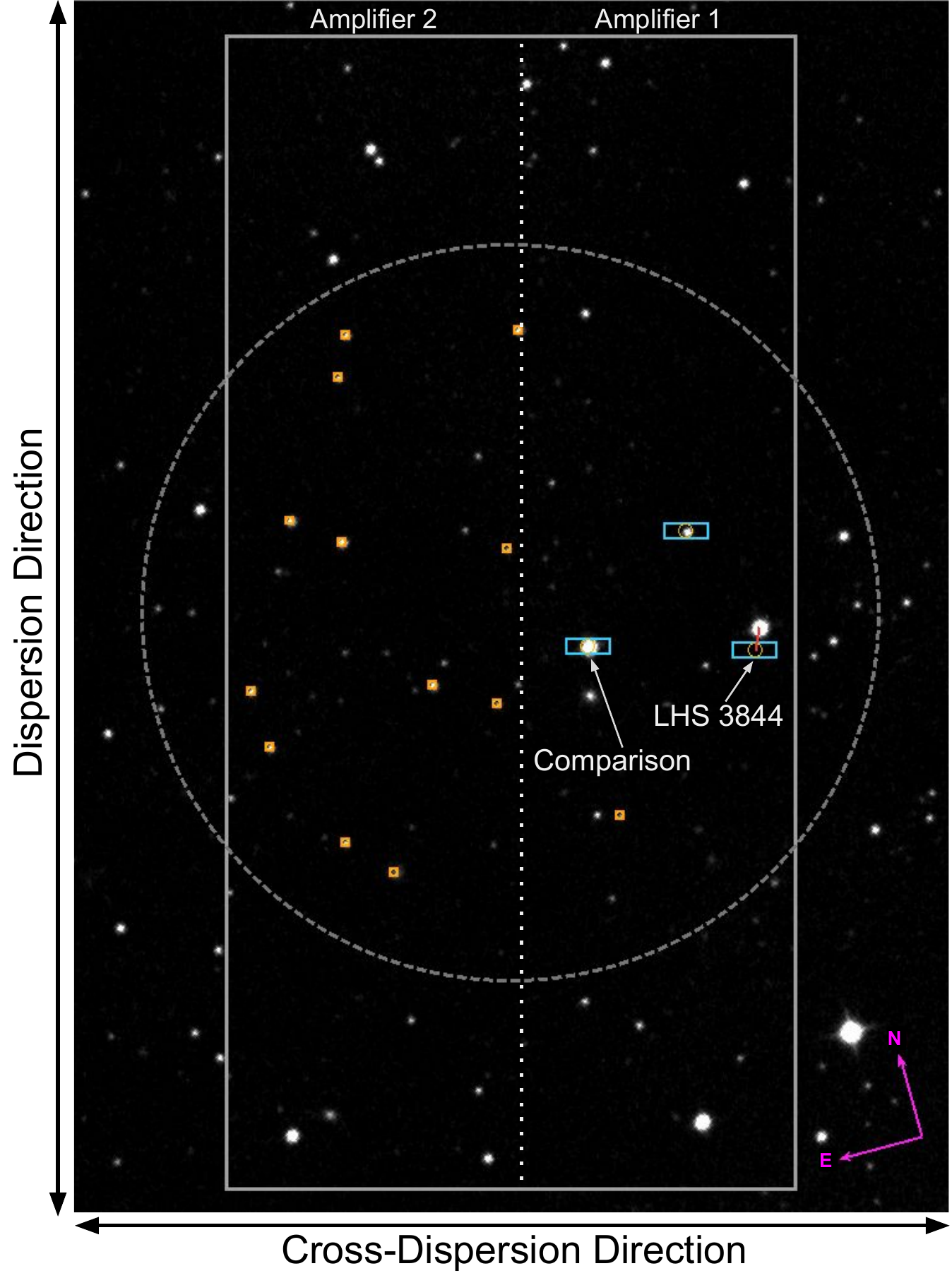}
\caption{Image of the LHS 3844 field from the Digitized Sky Survey (DSS), which is available in \texttt{SAOImagesDS9}. The dashed circle indicates the field of view of the LDSS3C multi-object spectrograph, with the solid rectangle showing the extent of the LDSS3C detector. LDSS3C has two amplifiers, which read out separately. These are labeled at the top of the detector rectangle, with the vertical dotted line indicating the split on the detector. Light blue rectangles indicate slits for LHS 3844 and comparison stars. A red line shows the path of LHS 3844 from 2000, when DSS observed this part of the sky, to 2019, when our observations were made. Only one comparison star is bright enough to use in the analysis. Orange squares indicate alignment star holes.}
\label{fig:onsky}
\end{figure}

\begin{deluxetable}{c|cc}
\tablecaption{Stars used in this work \label{tab:stars}}
\tablewidth{0pt}
\tablehead{
\colhead{} & \colhead{Target} & \colhead{Comparison}\\
}
\startdata
Name & LHS 3844 & 2MASS 22421963-6909508\\
RA  &  22:41:58.12 & 22:42:19.64 \\
Dec & -69:10:08.32 & -69:09:50.92\\
$V$ (mag)   & 15.24  & 12.557\\
$TESS$ (mag) & 11.9238 & 12.015\\
$J$ (mag)    & 10.046   & 11.438\\
\enddata
\tablecomments{Values are from TESS Input Catalog version 8.}
\end{deluxetable}

The field-of-view of LDSS3C is 8.3$'$ in diameter. The field-of-view is further cropped to 6.4$'$ in the cross-dispersion direction when translated onto the CCD. As of June 2019, amplifier two on the LDSS3C detector, which corresponds to the left side of the CCD chip, suffers from poor electronic connections. After our first observation we re-made our observing masks so as to avoid putting any science spectra on that part of the chip, and instead used it only for alignment stars (Figure~\ref{fig:onsky}). This further curtailing of the LDSS3C field-of-view means that we were limited in our choice of comparison stars. In the reduced field-of-view of LDSS3C we were able to observe two comparison stars, only one of which, 2MASS 22421963-6909508, is bright enough to calibrate LHS 3844 (Table~\ref{tab:stars}).

\section{Data Extraction \& Analysis} \label{sec:extractanalysis}

To extract and analyze our data we use two custom pipelines developed for ground-based multi-object spectroscopy. The extraction pipeline, \texttt{mosasaurus}\footnote{\href{http://github.com/zkbt/mosasaurus}{github.com/zkbt/mosasaurus}}, turns the raw images collected by Magellan II/LDSS3C into wavelength-calibrated spectra for LHS 3844 and the comparison star \citep{Diamond-Lowe2020a}. We subtract biases and darks from every science image and cut out extraction rectangles around the LHS 3844 and comparison star spectra. We divide the extraction rectangles by their associated flats and designate a fixed-width aperture around the stellar trace. We sum the flux in the cross-dispersion direction to create a spectrum, and then subtract off the estimated sky background. We use He, Ne, and Ar arcs taken during daytime calibrations to perform an initial wavelength calibration, and we use a cross-correlation function to adjust all spectra in a data set onto a common wavelength grid. 

\begin{figure}
\includegraphics[width=0.485\textwidth]{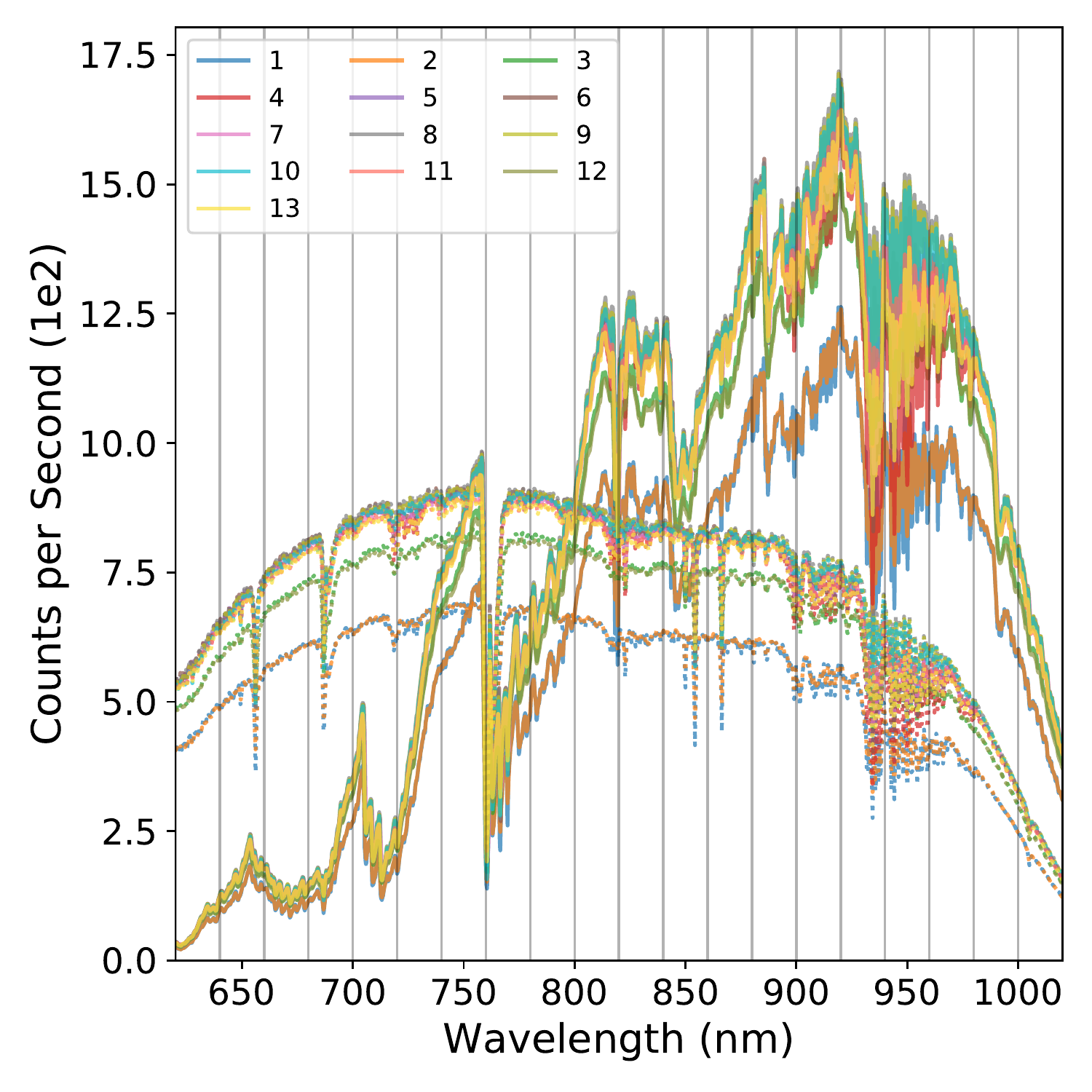}
\caption{Spectra of LHS 3844 (solid lines) and the comparison star (dotted lines) from each of the 13 data sets used in this work. The spectra represent the total counts from each star measured over the course of each observing night, normalized by the total exposure time (exposure time $\times$ number of exposures). Vertical gray lines are the edges of the 20 nm spectrophotometric bands.}
\label{fig:spectrabinned}
\end{figure}

We present spectra of LHS 3844 and the comparison star for each of the 13 data sets (Figure~\ref{fig:spectrabinned}). Unlike LHS 3844, the comparison star is not an M dwarf; it is brighter than LHS 3844 at optical wavelengths, but becomes dimmer than LHS 3844 at wavelengths redder than 800 nm. This means that in wavelength bins redder than 800 nm we are photon-limited by the comparison star rather than by LHS 3844.

The analysis pipeline, \texttt{decorrasaurus}\footnote{\href{https://github.com/hdiamondlowe/decorrasaurus/releases/tag/v2.0}{github.com/hdiamondlowe/decorrasaurus}} takes the wavelength-calibrated spectra and creates decorrelated, spectroscopic light curves that can be used to construct a transmission spectrum \citep{Diamond-Lowe2020a}. We divide the LHS 3844 time series by the comparison star time series to create our light curves. We then process this light curve to remove remaining correlated noise while simultaneously fitting a transit model. For this work we use the Gaussian process (GP) regression capabilities of \texttt{decorrasaurus}.

For this project we make some changes to our process of choosing the best input vectors to use in the construction of GP covariance matrix \citep[Section 4.2.1 of ][]{Diamond-Lowe2020a}. We still use the white light curve to test combinations of input vectors, and then select the optimal combination using the Bayesian evidence derived from the dynamic nested sampler, \texttt{dynesty} \citep{Speagle2020}. We then find the optimal input vectors for two 20 nm spectroscopic bands, one towards the blue end of the spectrum (710-730 nm), and one towards the red end of the spectrum (930-950 nm). If additional input vectors are needed in either of the spectroscopic bands, we include those in the fit. We also fix the priors for all length scales associated with the input vectors to be in the range of [$ln(10^{-2}), ln(10^{5})$]. This ensures equal prior volumes for all length scales. As before, we use the same input vectors for the white light and spectroscopic fits for a given data set. The input vectors we use in each data set are provided in Table~\ref{tab:sysparams}. More detailed explanations of the physical meanings of these vectors are found in Table 3 of \citet{Diamond-Lowe2020a}.

\definecolor{Gray}{gray}{0.85}
\newcolumntype{g}{>{\columncolor{Gray}[1.5mm][1.5mm]}c}
\begin{deluxetable*}{lgcgcgcgcgcgcg}
\tablecaption{Gaussian process input vectors\label{tab:sysparams}}
\tablewidth{0pt}
\tablehead{
\colhead{Input} &\multicolumn{13}{c}{Data set number}\\
\cline{2-14}
\colhead{Vectors} & \colhead{1} & \colhead{2} & \colhead{3} & \colhead{4} & \colhead{5} &\colhead{6} & \colhead{7} & \colhead{8} & \colhead{9} & \colhead{10} & \colhead{11} & \colhead{12} & \colhead{13}
}
\startdata
\textbf{airmass}        & \checkmark & \checkmark &            & \checkmark & \checkmark & \checkmark & \checkmark & \checkmark & \checkmark & \checkmark & \checkmark & \checkmark & \checkmark \\
\textbf{rotation angle} &            &            &            &            &            &            &            &            &            &            &            & \checkmark &            \\
\textbf{centroid}       &            &            &            &            & \checkmark &            &            &            &            &            &            &            &            \\
\textbf{width}          & \checkmark & \checkmark &            &            & \checkmark & \checkmark &            &            & \checkmark &            & \checkmark &            &            \\
\textbf{peak}           & \checkmark & \checkmark & \checkmark & \checkmark &            &            &            &            &            &            & \checkmark & \checkmark & \checkmark \\
\textbf{shift}          &            & \checkmark &            &            &            & \checkmark & \checkmark &            &            & \checkmark &            &            &            \\
\textbf{stretch}        &            &            &            &            &            &            &            &            &            &            &            &            &            \\
\textbf{time}           & \checkmark &            &            &            & \checkmark &            &            & \checkmark &            & \checkmark & \checkmark &            & \checkmark \\
\enddata
\tablecomments{A more detailed explanation of the input vectors can by found in \citep{Diamond-Lowe2020a}.}
\end{deluxetable*}

\subsection{White light curves} \label{subsec:whitelc}

For each of our 13 data sets, we perform a white light curve fit from 620-1020 nm. LHS 3844b has a short circularization timescale, and \citet{Kreidberg2019} find the secondary eclipse of LHS 3844b at phase 0.5; we therefore fix the orbital eccentricity to zero. We fix the period $P$ and transit ephemeris $T_0$ to the values revised by \citet{Kreidberg2019}. Note that there is an 0.5 JD error in $T_0$ given in \citet{Kreidberg2019}; we use $T_0$=2458325.72559 $BJD_{TDB}$ (L. Kreidberg, priv.\ comm.). We fit for the inclination $i$ and scaled semi-major axis $a/R_s$. 

Similar to the analysis performed in \citet{Diamond-Lowe2020a}, we employ a logarithmic limb-darkening law, and use the Limb-Darkening Toolkit \citep[\texttt{LDTk}][]{Parviainen2015} to produce the coefficients. We re-parameterize these coefficients according to \citet{Espinoza2017}, and fit for them. In this analysis we place Gaussian priors on the re-parameterized limb-darkening coefficients $q_0$ and $q_1$, with the mean and width of the Gaussian set by the mean and 5$\times$ the uncertainty returned by \texttt{LDTk}. The wide Gaussian priors on these parameters take into account our limited understanding of how stars darken towards their limbs as well as the additional factors that are not accounted for by \texttt{LDTk} such as the wavelength sensitivity of LDSS3C, Magellan Clay, and Earth's atmosphere. By not using a flat prior we cut down on parameter space to explore, thereby decreasing the computational time to perform the fit.

Before performing a full exploration of the parameter space, we compute the predictive distribution of the GP model on the white light curve. While this fit is not optimal, it allows us to clip outlying data points that fall outside 5$\times$ the mean absolute deviation of the residuals. We clip as few as 0\% and as many as 2.6\% of the white light curve data points. Table~\ref{tab:whitelc} presents the priors we place on the white light curve transit parameters, and the derived parameters for each data set. We also provide the RMS of each white light curve fit compared to its GP and transit model.

\begin{figure}
\includegraphics[width=0.47\textwidth]{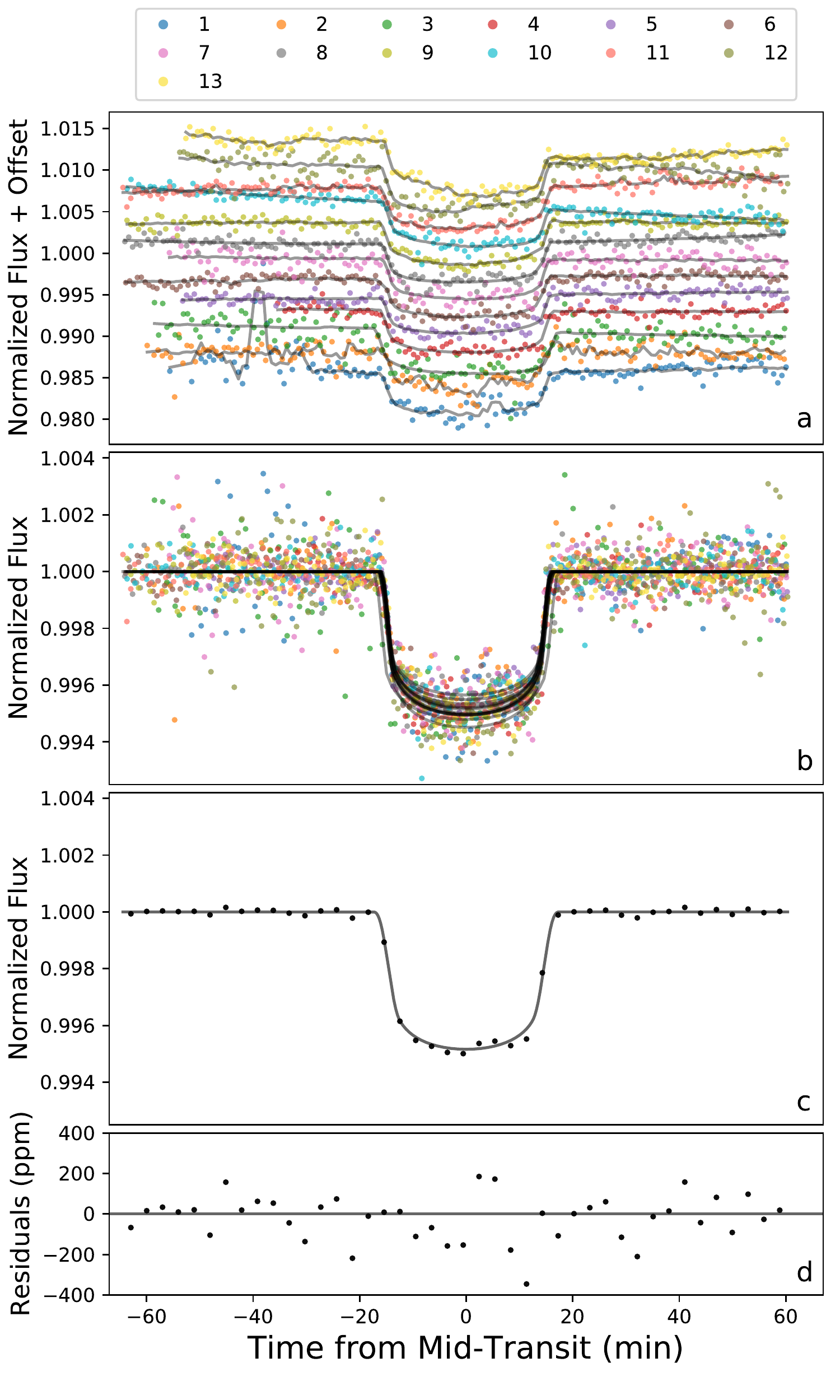}
\caption{\textit{Panel a}: Raw white light curves for the 13 data sets included in this work. Data set numbers correspond to those in Table~\ref{tab:obs}. Over-plotted grey lines are models computed by the Gaussian process regression, which includes the transit model as the mean function. We fit each of the data sets separately. \textit{Panel b}: White light curves of the 13 data sets with the Gaussian process model removed. The grey lines are the derived transit model for each data set. \textit{Panel c}: Data from all 13 data sets are combined and binned down to 3-minute time bins. The grey transit model is constructed using the inverse-variance weighted mean of the 13 derived values of $R_p/R_s$, $i$, $a/R_s$, and the limb-darkening coefficients $q_0$ and $q_1$. The model is sampled at high-cadence and then smoothed with a 3-minute box-car kernel. \textit{Panel d}: Residuals of Panel c. With 3-minute time bins we achieve an RMS of 112 ppm; binning to 10-minutes gives an RMS of 65 ppm.}
\label{fig:whitelc}
\end{figure}

We present the raw white light curves in Panel a of Figure~\ref{fig:whitelc}, along with the individual models over-plotted. We do not see obvious evidence for spot-crossing events in any transit. Some particularly noisy transits, such as in Data Sets 1 and 2, are visibly correlated with poor visibility markers, such as the ``width'' input vector (Table~\ref{tab:sysparams}), which is a rough proxy for the seeing. We analyze each white light curve individually, and then take the inverse-variance weighted mean of their planet-to-star radius ratios $R_p/R_s$, inclinations $i$, scaled orbital distances $a/R_s$, and limb-darkening coefficients $q_0$ and $q_1$ to create a combined transit model.

In Panel b we remove the noise component of each model so that just the transit models are left. We plot each of the 13 transit models, and it is apparent that there is some dispersion in model parameters across the data sets. In Panel c we bin the data to 3-minute time bins and plot a high-cadence version of the combined transit model, smoothed with a 3-minute box-car kernel. To determine the RMS for the combined data sets we remove the correlated noise component of the GP model from each respective data set, and then combine the 13 light curves together. We calculate the RMS by comparing the combined data sets to the combined transit model. In 3-minute time bins we achieve an RMS precision of 112 ppm. Binning to 10-minute bins gives an RMS of 65 ppm, but past that the residuals do not bin down as predicted for $1/\sqrt{N}$ where $N$ is the number of data points in a time bin.

\begin{splitdeluxetable*}{lc|cccccBlc|ccccBlc|cccc}
\tablecaption{White light curve transit parameters and priors\label{tab:whitelc}}
\tablewidth{0pt}
\tablehead{
\multirow{2}{*}{Parameter} & \multirow{2}{*}{Prior} &\multicolumn{5}{c}{Data set numbers} & \multirow{2}{*}{Parameter} & \multirow{2}{*}{Prior} &\multicolumn{4}{c}{Data set numbers} & \multirow{2}{*}{Parameter} & \multirow{2}{*}{Prior} & \multicolumn{4}{c}{Data set numbers}\\
& & \colhead{1} & \colhead{2} & \colhead{3} & \colhead{4} & \colhead{5} &  &  & \colhead{6} & \colhead{7} & \colhead{8} & \colhead{9} & & & \colhead{10} & \colhead{11} & \colhead{12} & \colhead{13}\\
}
\startdata
$\delta t$ (days) $^a$ & $\mathcal{U}$(-0.003, 0.003) $^b$  & 0.00125 $\pm$ 0.00022 & 0.00140 $\pm$ 0.00019 & 0.00155 $\pm$ 0.00029 & 0.00115 $\pm$ 0.00012 & 0.00131 $\pm$ 0.00010 & $\delta t$ (days) & $\mathcal{U}$(-0.003, 0.003) & 0.00150 $\pm$ 0.00011 & 0.00139 $\pm$ 0.00020 & 0.00134 $\pm$ 0.00011 & 0.00130 $\pm$ 0.00013 & $\delta t$ (days) & $\mathcal{U}$(-0.003, 0.003) & 0.00139 $\pm$ 0.00011 & 0.00169 $\pm$ 0.00011 & 0.00127 $\pm$ 0.00017 & 0.00119 $\pm$ 0.00009 \\ 
$R_p/R_s$ & $\mathcal{U}(0.055, 0.075)$ $^b$ & 0.0658 $\pm$ 0.0024 & 0.0623 $\pm$ 0.0016 & 0.0670 $\pm$ 0.0018 & 0.0659 $\pm$ 0.0011 & 0.0611 $\pm$ 0.0010 & $R_p/R_s$ & $\mathcal{U}(0.055, 0.075)$ & 0.0646 $\pm$ 0.0011 & 0.0641 $\pm$ 0.0017 & 0.0636 $\pm$ 0.0011 & 0.0659 $\pm$ 0.0010 & $R_p/R_s$ & $\mathcal{U}(0.055, 0.075)$ & 0.0642 $\pm$ 0.0011 & 0.0658 $\pm$ 0.0011 & 0.0686 $\pm$ 0.0017 & 0.0658 $\pm$ 0.0021 \\ 
$a/R_s$ & $\mathcal{U}(5, 10)$ $^b$ & $6.84^{+0.29}_{-0.41}$ & $6.90^{+0.27}_{-0.41}$ & $6.37^{+0.30}_{-0.35}$ & $6.79^{+0.23}_{-0.41}$ & $6.95^{+0.17}_{-0.35}$ & $a/R_s$ & $\mathcal{U}(5, 10)$ & $6.81^{+0.31}_{-0.42}$ & $6.97^{+0.27}_{-0.44}$ & $6.90^{+0.23}_{-0.40}$ & $6.45^{+0.40}_{-0.30}$ & $a/R_s$ & $\mathcal{U}(5, 10)$ & $7.03^{+0.21}_{-0.40}$ & $6.43^{+0.41}_{-0.28}$ & $6.97^{+0.22}_{-0.39}$ & $7.07^{+0.15}_{-0.33}$ \\ 
$i$ ($^{\circ}$) & $\mathcal{U}(85, 90)$ $^b$ & $87.7^{+1.5}_{-1.6}$ & $87.7^{+1.4}_{-1.5}$ & $87.5^{+1.5}_{-1.5}$ & $87.7^{+1.4}_{-1.5}$ & $88.1^{+1.1}_{-1.4}$ & $i$ ($^{\circ}$) & $\mathcal{U}(85, 90)$ & $87.2^{+1.6}_{-1.4}$ & $87.8^{+1.4}_{-1.6}$ & $87.7^{+1.4}_{-1.5}$ & $86.5^{+1.8}_{-1.0}$ & $i$ ($^{\circ}$) & $\mathcal{U}(85, 90)$ & $88.0^{+1.2}_{-1.5}$ & $86.3^{+1.7}_{-0.9}$ & $88.0^{+1.2}_{-1.5}$ & $88.3^{+1.0}_{-1.4}$ \\ 
$q_0$ & $\mathcal{N}(0.38, 0.13)$ $^b$ & 0.38 $\pm$ 0.12 & 0.37 $\pm$ 0.12 & 0.37 $\pm$ 0.12 & 0.33 $\pm$ 0.12 & 0.36 $\pm$ 0.11 & $q_0$ & $\mathcal{N}(0.38, 0.13)$ & 0.40 $\pm$ 0.11 & 0.37 $\pm$ 0.12 & 0.30 $\pm$ 0.15 & 0.46 $\pm$ 0.13 & $q_0$ & $\mathcal{N}(0.38, 0.13)$ & 0.38 $\pm$ 0.14 & 0.40 $\pm$ 0.12 & 0.37 $\pm$ 0.11 & 0.42 $\pm$ 0.12 \\ 
$q_1$ & $\mathcal{N}(0.475, 0.057)$ $^b$ & 0.461 $\pm$ 0.051 & 0.475 $\pm$ 0.051 & 0.474 $\pm$ 0.049 & 0.490 $\pm$ 0.045 & 0.481 $\pm$ 0.043 & $q_1$ & $\mathcal{N}(0.475, 0.057)$ & 0.456 $\pm$ 0.048 & 0.481 $\pm$ 0.049 & 0.500 $\pm$ 0.058 & 0.424 $\pm$ 0.054 & $q_1$ & $\mathcal{N}(0.475, 0.057)$ & 0.474 $\pm$ 0.059 & 0.443 $\pm$ 0.051 & 0.478 $\pm$ 0.048 & 0.473 $\pm$ 0.048 \\ 
 \multicolumn{2}{c}{RMS (ppm) $^c$}   & 1034 & 905 & 1230 & 577 & 548 &  \multicolumn{2}{c}{RMS (ppm)} & 514 & 989 & 549 & 466 &  \multicolumn{2}{c}{RMS (ppm)} & 597 & 473 & 1101 & 537 \\ 
\enddata
\tablecomments{\\
\tsup{a} $\delta t$ is the difference between the predicted time of mid transit and the derived time-of mid transit from fitting each white light curve. The predicted time of mid-transit is calculated as $t_0 = T_0 + nP$ where $T_0 = 2458325.72559\ BJD_{TDB}$, $P = 0.46292792$ days, and $n$ is the transit number provided in Table~\ref{tab:obs}.\\
\tsup{b} $\mathcal{U}$ denotes a uniform prior; $\mathcal{N}$ denotes a Gaussian prior.\\
\tsup{c} The RMS values in the bottom row refer to the RMS of the white light curve residuals in each individual data set compared to the model. The RMS of the combined data sets is discussed in Figure~\ref{fig:whitelc}.}
\end{splitdeluxetable*}

\subsection{Spectroscopic light curves}\label{subsec:spectro}

\begin{figure*}
\includegraphics[width=\textwidth]{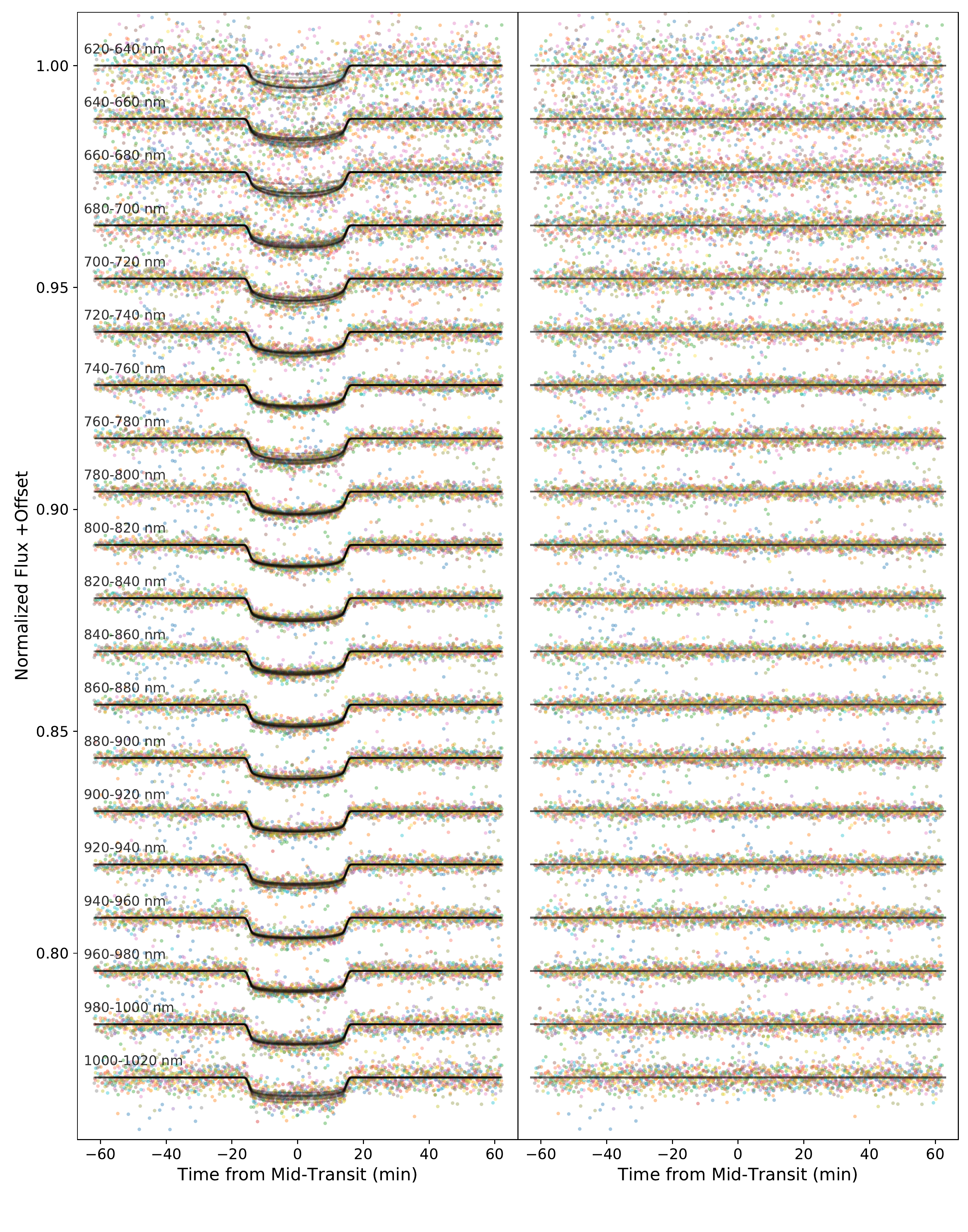}
\vspace{-25pt}
\caption{\textit{Left}: Band-integrated spectroscopic light curves with GP models removed for each of the 13 data sets. The transit model (GP mean function) is plotted in grey for each data set. Spectroscopic bands are offset for clarity. \textit{Right}: Residuals of all 13 data sets, in each spectroscopic band, also offset for clarity. The spectroscopic transit depths and uncertainties are provided in Table~\ref{tab:transitdepths} and Figure~\ref{fig:transmissionhist}.}
\label{fig:wavelc}
\end{figure*}

We split the white light curves into 20 spectroscopic bands of 20 nm each to create our band-integrated spectroscopic light curves. In each band, and for each data set, we fix the times of mid-transit to the values derived from the white light curves (Table~\ref{tab:whitelc}). We fix the values of $i$ and $a/R_s$ to the inverse-variance weighted means of the white light curve fits in the 13 data sets; $i=87.9^{\circ}$, $a/R_s=6.83$. We then fit for the planet-to-star radius ratio $R_p/R_s$ and the re-parameterized logarithmic limb-darkening coefficients $q_0$ and $q_1$. We present the resulting light curves with the GP noise component removed in Figure~\ref{fig:wavelc}, and list the transit depths $(R_p/R_s)^2$ for each data set in Table~\ref{tab:transitdepths}. In Table~\ref{tab:transitdepths} we also provide the inverse-variance weighted mean of the transit depths in each spectroscopic band, the RMS of all 13 combined data sets (without any time binning), and how close we get to the calculated photon noise in each band, averaged across the 13 data sets. We calculate the photon noise in each data set by summing the raw photoelectron counts $R$ and estimated sky background counts $S$ in the spectroscopic band for both the target and comparison star. The calculated photon noise for the target star is $\sigma_T = \sqrt{R_T + S_T}/R_T$; we calculate the same for the comparison star $\sigma_C$. We take the calculated photon noise for the combined target and comparison star observations as $\sqrt{\sigma_T^2 + \sigma_C^2}$.

\begin{figure*}
\includegraphics[width=\textwidth]{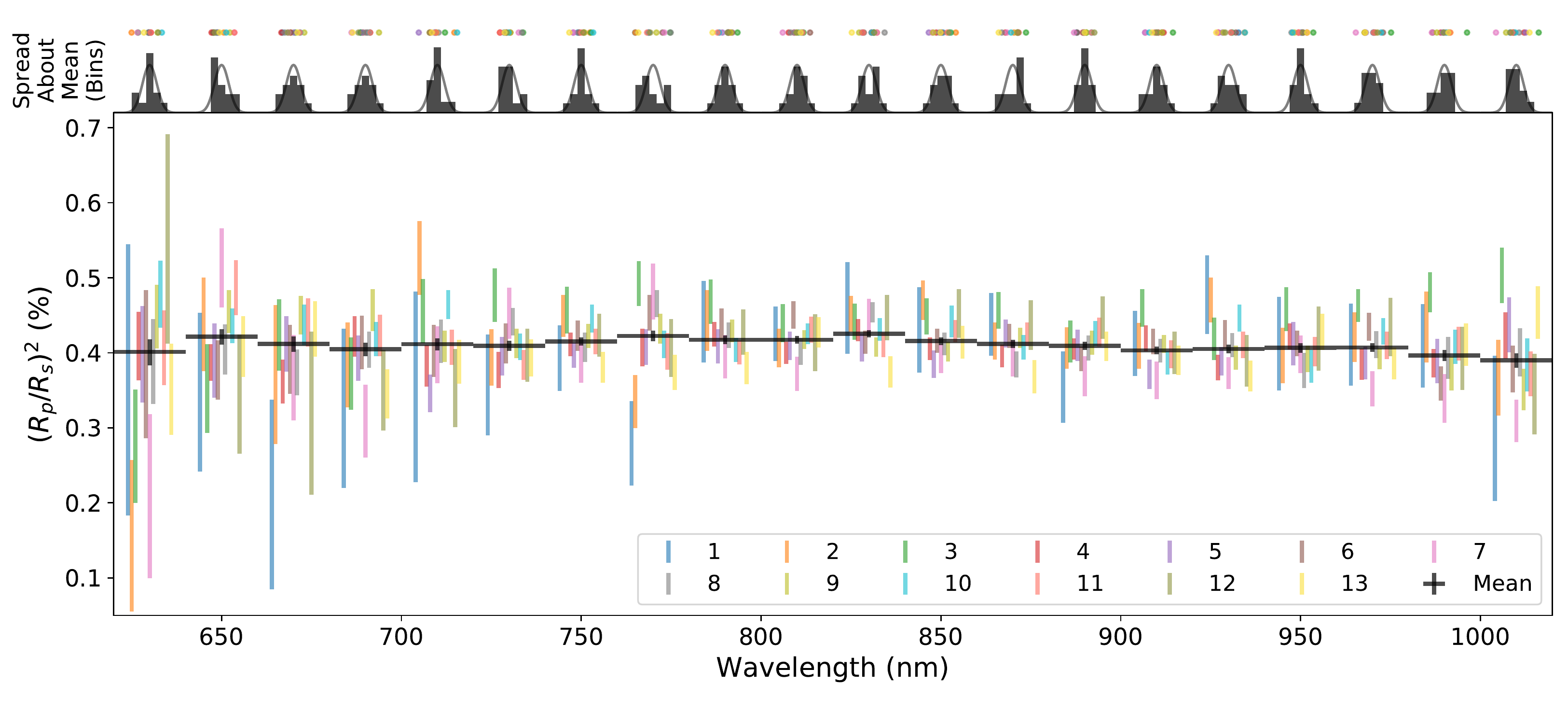}
\caption{Transmission spectra of LHS 3844b from each of the 13 data sets analyzed. Values of $(R_p/R_s)^2$ along with 1-$\sigma$ error bars are provided for each of the 13 data sets in each spectroscopic band. Transit depths are offset in the x-axis for clarity. Black points with 1-$\sigma$ error bars are the inverse-variance weighted mean transit depths in each band. Horizontal bars on the black points denote the span of each 20-nm spectroscopic band. On top are the spread of each transit depth about the inverse-variance weighted mean in each band; i.e., $(D_i - \bar D)/\sigma_{D_i}$, compared to Gaussian distributions. The dots above the histograms show where this value lands for each data set. We do not find that any data sets are systematically offset from the others.}
\label{fig:transmissionhist}
\end{figure*}

We present transmission spectra from each data set, fit individually, along with the inverse-variance weighted mean of the transit depths in Figure~\ref{fig:transmissionhist}. The histograms above each spectroscopic band show the spread in transit depths according to $\left(D_i - \bar D\right)/\sigma_{D_i}$, where $D_i$ and $\sigma_{D_i}$ are the transit depth and 1-$\sigma$ transit depth uncertainty for each data set $i$, and $\bar D$ is the inverse-variance weighted mean transit depth in that band. The dots on top of the histograms show where each data set is in relation to the others. We do not find that any data set is consistently lower or higher than the mean. Increased transit depth uncertainties in the bluest spectroscopic bands are most likely due to low levels of flux from the host M dwarf LHS 3844, while increased transit depth uncertainties in the reddest spectroscopic bands likely arise from a decrease in stellar flux due to absorption by telluric water vapor, as well as decreasing throughput and quantum efficiency of LDSS3C and its detector.

\definecolor{Gray}{gray}{0.85}
\begin{splitdeluxetable*}{c|ccccccccBc|cccccggg}
\tablecaption{Spectroscopic transit depths\label{tab:transitdepths}}
\tablewidth{0pt}
\tablehead{
\rowcolor{white}\colhead{Wavelength} & \multicolumn{8}{c}{Transit depths by data set (\%)} & \colhead{Wavelength} & \multicolumn{5}{c}{Transit depths by data set (\%)} & \colhead{Mean} & \multirow{2}{*}{RMS} & \colhead{$\times$ Exp.} \\
\rowcolor{white} \colhead{(nm)}& \colhead{1} & \colhead{2} & \colhead{3} & \colhead{4} & \colhead{5} &  \colhead{6} & \colhead{7} & \colhead{8} & \colhead{(nm)} & \colhead{9} & \colhead{10} & \colhead{11} & \colhead{12} & \colhead{13} & \colhead{(\%)} & \colhead{(ppm)} & \colhead{Noise}\\
}
\startdata
620-640 & 0.383 $\pm$ 0.181 & 0.147 $\pm$ 0.101 & 0.280 $\pm$ 0.076 & 0.409 $\pm$ 0.046 & 0.401 $\pm$ 0.064 & 0.390 $\pm$ 0.099 & 0.218 $\pm$ 0.110 & 0.390 $\pm$ 0.057 &  620-640 & 0.449 $\pm$ 0.043 & 0.478 $\pm$ 0.045 & 0.409 $\pm$ 0.050 & 0.566 $\pm$ 0.140 & 0.353 $\pm$ 0.061 & 0.4012 $\pm$ 0.0174 & 5377 & 1.79 \\ 
640-660 & 0.349 $\pm$ 0.106 & 0.441 $\pm$ 0.063 & 0.352 $\pm$ 0.059 & 0.387 $\pm$ 0.025 & 0.392 $\pm$ 0.050 & 0.378 $\pm$ 0.040 & 0.515 $\pm$ 0.053 & 0.405 $\pm$ 0.033 &  640-660 & 0.455 $\pm$ 0.029 & 0.437 $\pm$ 0.023 & 0.487 $\pm$ 0.037 & 0.349 $\pm$ 0.080 & 0.410 $\pm$ 0.041 & 0.4215 $\pm$ 0.0105 & 2702 & 1.71 \\ 
660-680 & 0.217 $\pm$ 0.126 & 0.376 $\pm$ 0.093 & 0.424 $\pm$ 0.048 & 0.362 $\pm$ 0.029 & 0.413 $\pm$ 0.037 & 0.393 $\pm$ 0.046 & 0.370 $\pm$ 0.057 & 0.375 $\pm$ 0.030 &  660-680 & 0.450 $\pm$ 0.026 & 0.437 $\pm$ 0.027 & 0.440 $\pm$ 0.032 & 0.325 $\pm$ 0.109 & 0.432 $\pm$ 0.037 & 0.4118 $\pm$ 0.0105 & 3188 & 1.64 \\ 
680-700 & 0.331 $\pm$ 0.106 & 0.386 $\pm$ 0.056 & 0.374 $\pm$ 0.048 & 0.422 $\pm$ 0.027 & 0.394 $\pm$ 0.030 & 0.414 $\pm$ 0.036 & 0.311 $\pm$ 0.049 & 0.406 $\pm$ 0.024 &  680-700 & 0.457 $\pm$ 0.028 & 0.418 $\pm$ 0.023 & 0.423 $\pm$ 0.028 & 0.353 $\pm$ 0.054 & 0.346 $\pm$ 0.033 & 0.4048 $\pm$ 0.0091 & 2444 & 1.67 \\ 
700-720 & 0.361 $\pm$ 0.127 & 0.528 $\pm$ 0.049 & 0.457 $\pm$ 0.043 & 0.383 $\pm$ 0.028 & 0.347 $\pm$ 0.025 & 0.403 $\pm$ 0.034 & 0.398 $\pm$ 0.038 & 0.416 $\pm$ 0.029 &  700-720 & 0.408 $\pm$ 0.021 & 0.465 $\pm$ 0.019 & 0.407 $\pm$ 0.023 & 0.355 $\pm$ 0.052 & 0.389 $\pm$ 0.029 & 0.4115 $\pm$ 0.0082 & 2056 & 1.66 \\ 
720-740 & 0.361 $\pm$ 0.067 & 0.394 $\pm$ 0.038 & 0.477 $\pm$ 0.035 & 0.378 $\pm$ 0.024 & 0.396 $\pm$ 0.026 & 0.413 $\pm$ 0.026 & 0.453 $\pm$ 0.034 & 0.442 $\pm$ 0.019 &  720-740 & 0.413 $\pm$ 0.020 & 0.408 $\pm$ 0.018 & 0.384 $\pm$ 0.020 & 0.396 $\pm$ 0.035 & 0.393 $\pm$ 0.025 & 0.4093 $\pm$ 0.0069 & 1650 & 1.73 \\ 
740-760 & 0.394 $\pm$ 0.044 & 0.449 $\pm$ 0.028 & 0.457 $\pm$ 0.031 & 0.411 $\pm$ 0.016 & 0.396 $\pm$ 0.017 & 0.422 $\pm$ 0.020 & 0.387 $\pm$ 0.026 & 0.408 $\pm$ 0.020 &  740-760 & 0.423 $\pm$ 0.016 & 0.446 $\pm$ 0.019 & 0.415 $\pm$ 0.017 & 0.423 $\pm$ 0.029 & 0.381 $\pm$ 0.021 & 0.4151 $\pm$ 0.0057 & 1214 & 1.57 \\ 
760-780 & 0.282 $\pm$ 0.056 & 0.336 $\pm$ 0.035 & 0.493 $\pm$ 0.030 & 0.408 $\pm$ 0.025 & 0.408 $\pm$ 0.024 & 0.454 $\pm$ 0.024 & 0.480 $\pm$ 0.037 & 0.466 $\pm$ 0.018 &  760-780 & 0.432 $\pm$ 0.020 & 0.410 $\pm$ 0.019 & 0.402 $\pm$ 0.024 & 0.407 $\pm$ 0.039 & 0.374 $\pm$ 0.023 & 0.4226 $\pm$ 0.0069 & 1526 & 1.59 \\ 
780-800 & 0.443 $\pm$ 0.054 & 0.443 $\pm$ 0.040 & 0.469 $\pm$ 0.030 & 0.425 $\pm$ 0.016 & 0.409 $\pm$ 0.023 & 0.438 $\pm$ 0.021 & 0.396 $\pm$ 0.029 & 0.424 $\pm$ 0.017 &  780-800 & 0.429 $\pm$ 0.015 & 0.404 $\pm$ 0.016 & 0.401 $\pm$ 0.016 & 0.428 $\pm$ 0.030 & 0.380 $\pm$ 0.021 & 0.4174 $\pm$ 0.0058 & 1383 & 1.64 \\ 
800-820 & 0.426 $\pm$ 0.036 & 0.407 $\pm$ 0.026 & 0.440 $\pm$ 0.025 & 0.400 $\pm$ 0.015 & 0.410 $\pm$ 0.019 & 0.450 $\pm$ 0.018 & 0.372 $\pm$ 0.023 & 0.401 $\pm$ 0.018 &  800-820 & 0.418 $\pm$ 0.013 & 0.426 $\pm$ 0.014 & 0.431 $\pm$ 0.017 & 0.413 $\pm$ 0.038 & 0.428 $\pm$ 0.020 & 0.4173 $\pm$ 0.0052 & 1125 & 1.56 \\ 
820-840 & 0.464 $\pm$ 0.061 & 0.452 $\pm$ 0.024 & 0.442 $\pm$ 0.024 & 0.431 $\pm$ 0.015 & 0.407 $\pm$ 0.019 & 0.414 $\pm$ 0.016 & 0.450 $\pm$ 0.022 & 0.454 $\pm$ 0.013 &  820-840 & 0.408 $\pm$ 0.013 & 0.431 $\pm$ 0.015 & 0.411 $\pm$ 0.016 & 0.447 $\pm$ 0.030 & 0.375 $\pm$ 0.021 & 0.4254 $\pm$ 0.0050 & 1099 & 1.50 \\ 
840-860 & 0.432 $\pm$ 0.057 & 0.470 $\pm$ 0.026 & 0.449 $\pm$ 0.024 & 0.406 $\pm$ 0.015 & 0.385 $\pm$ 0.018 & 0.416 $\pm$ 0.016 & 0.392 $\pm$ 0.019 & 0.412 $\pm$ 0.015 &  840-860 & 0.401 $\pm$ 0.013 & 0.441 $\pm$ 0.021 & 0.429 $\pm$ 0.015 & 0.452 $\pm$ 0.032 & 0.415 $\pm$ 0.022 & 0.4156 $\pm$ 0.0052 & 1185 & 1.53 \\ 
860-880 & 0.439 $\pm$ 0.042 & 0.416 $\pm$ 0.025 & 0.457 $\pm$ 0.025 & 0.396 $\pm$ 0.015 & 0.425 $\pm$ 0.019 & 0.422 $\pm$ 0.016 & 0.392 $\pm$ 0.024 & 0.385 $\pm$ 0.017 &  860-880 & 0.420 $\pm$ 0.014 & 0.408 $\pm$ 0.016 & 0.425 $\pm$ 0.016 & 0.436 $\pm$ 0.033 & 0.368 $\pm$ 0.022 & 0.4118 $\pm$ 0.0053 & 1157 & 1.57 \\ 
880-900 & 0.356 $\pm$ 0.048 & 0.406 $\pm$ 0.028 & 0.416 $\pm$ 0.028 & 0.406 $\pm$ 0.014 & 0.410 $\pm$ 0.021 & 0.392 $\pm$ 0.016 & 0.369 $\pm$ 0.027 & 0.407 $\pm$ 0.017 &  880-900 & 0.413 $\pm$ 0.016 & 0.425 $\pm$ 0.017 & 0.428 $\pm$ 0.019 & 0.447 $\pm$ 0.028 & 0.409 $\pm$ 0.020 & 0.4093 $\pm$ 0.0055 & 1333 & 1.76 \\ 
900-920 & 0.413 $\pm$ 0.044 & 0.409 $\pm$ 0.031 & 0.460 $\pm$ 0.025 & 0.420 $\pm$ 0.017 & 0.371 $\pm$ 0.020 & 0.415 $\pm$ 0.017 & 0.364 $\pm$ 0.025 & 0.403 $\pm$ 0.016 &  900-920 & 0.412 $\pm$ 0.012 & 0.387 $\pm$ 0.016 & 0.398 $\pm$ 0.019 & 0.400 $\pm$ 0.029 & 0.390 $\pm$ 0.020 & 0.4031 $\pm$ 0.0053 & 1242 & 1.69 \\ 
920-940 & 0.478 $\pm$ 0.053 & 0.471 $\pm$ 0.030 & 0.420 $\pm$ 0.028 & 0.381 $\pm$ 0.017 & 0.389 $\pm$ 0.019 & 0.424 $\pm$ 0.019 & 0.374 $\pm$ 0.021 & 0.411 $\pm$ 0.016 &  920-940 & 0.394 $\pm$ 0.016 & 0.446 $\pm$ 0.018 & 0.411 $\pm$ 0.018 & 0.389 $\pm$ 0.034 & 0.369 $\pm$ 0.021 & 0.4051 $\pm$ 0.0057 & 1355 & 1.64 \\ 
940-960 & 0.412 $\pm$ 0.062 & 0.397 $\pm$ 0.037 & 0.459 $\pm$ 0.029 & 0.422 $\pm$ 0.018 & 0.412 $\pm$ 0.029 & 0.413 $\pm$ 0.017 & 0.398 $\pm$ 0.025 & 0.377 $\pm$ 0.023 &  940-960 & 0.392 $\pm$ 0.018 & 0.383 $\pm$ 0.022 & 0.401 $\pm$ 0.020 & 0.420 $\pm$ 0.043 & 0.429 $\pm$ 0.024 & 0.4065 $\pm$ 0.0065 & 1473 & 1.75 \\ 
960-980 & 0.413 $\pm$ 0.055 & 0.421 $\pm$ 0.033 & 0.464 $\pm$ 0.022 & 0.385 $\pm$ 0.021 & 0.387 $\pm$ 0.022 & 0.435 $\pm$ 0.019 & 0.353 $\pm$ 0.023 & 0.412 $\pm$ 0.018 &  960-980 & 0.392 $\pm$ 0.014 & 0.428 $\pm$ 0.018 & 0.410 $\pm$ 0.019 & 0.435 $\pm$ 0.039 & 0.386 $\pm$ 0.021 & 0.4071 $\pm$ 0.0058 & 1361 & 1.57 \\ 
980-1000 & 0.411 $\pm$ 0.056 & 0.436 $\pm$ 0.047 & 0.481 $\pm$ 0.027 & 0.387 $\pm$ 0.019 & 0.390 $\pm$ 0.029 & 0.359 $\pm$ 0.023 & 0.340 $\pm$ 0.032 & 0.394 $\pm$ 0.028 &  980-1000 & 0.373 $\pm$ 0.022 & 0.408 $\pm$ 0.023 & 0.412 $\pm$ 0.022 & 0.394 $\pm$ 0.042 & 0.412 $\pm$ 0.028 & 0.3965 $\pm$ 0.0075 & 1704 & 1.70 \\ 
1000-1020 & 0.306 $\pm$ 0.097 & 0.368 $\pm$ 0.051 & 0.504 $\pm$ 0.037 & 0.424 $\pm$ 0.031 & 0.438 $\pm$ 0.037 & 0.379 $\pm$ 0.031 & 0.310 $\pm$ 0.028 & 0.402 $\pm$ 0.032 &  1000-1020 & 0.351 $\pm$ 0.028 & 0.385 $\pm$ 0.035 & 0.373 $\pm$ 0.030 & 0.347 $\pm$ 0.054 & 0.454 $\pm$ 0.035 & 0.3901 $\pm$ 0.0097 & 2193 & 1.60 \\ 
\enddata
\tablecomments{The final three columns in grey provide the inverse-variance weighted mean across all 13 data sets for each spectroscopic band, along with the RMS of the 13 data sets combined in each band. The final column, $\times$ Expected Noise, describes how close, on average, we get to the calculated photon noise in each band.}
\end{splitdeluxetable*}

\section{Results \& Discussion} \label{sec:resultsdiscussion}

\subsection{LHS 3844b transmission spectrum compared to model transmission spectra}

We use the inverse-variance weighted mean of the derived transit depths in each of the 20-nm spectrophotometric bands as our combined observed transmission spectrum. We compare this observed transmission spectrum to model transmission spectra in order to address the atmosphere of LHS 3844b. We construct the model transmission spectra using two open-source codes: \texttt{HELIOS} \citep{Malik2017,Malik2019a,Malik2019b} and \texttt{Exo-Transmit} \citep{Miller-RicciKempton2012,Kempton2017}. 

With the \texttt{HELIOS} code we calculate the temperature-pressure profiles in radiative-convective equilibrium using the same numerical set-up (chemical abundances and opacities) as in \citet{Malik2019b}. We emulate planetary limb conditions in the 1-D radiative transfer model by setting the zenith angle for the incident stellar radiation to 80 degrees. We then feed these temperature profiles into \texttt{Exo-Transmit}. Like \citet{Kreidberg2019}, we assume an Earth-like bulk composition for LHS 3844b, which gives a surface gravity of 16 m/s$^2$ in our models. 

Both \texttt{HELIOS} and \texttt{Exo-Transmit} use a reference base of the atmosphere, such as a planet's rocky surface or an impenetrable cloud deck, where the optical depth $\tau\gg1$. \texttt{Exo-Transmit} allows the user to select as a free parameter the radius of the planet at the base of the atmosphere. For gas giant planets the definition of this parameter is somewhat arbitrary, but for a rocky planet like LHS 3844b, the base-of-atmosphere radius corresponds to that of the solid portion of the planet. We do not know \textit{a priori} where the solid-to-atmosphere transition is on LHS 3844b, so for each atmospheric case we vary the planet radius input to \texttt{Exo-Transmit} by 0.1\% until we find the lowest $\chi^2$ fit to the observed transmission spectrum.

\begin{figure*}
\includegraphics[width=\textwidth]{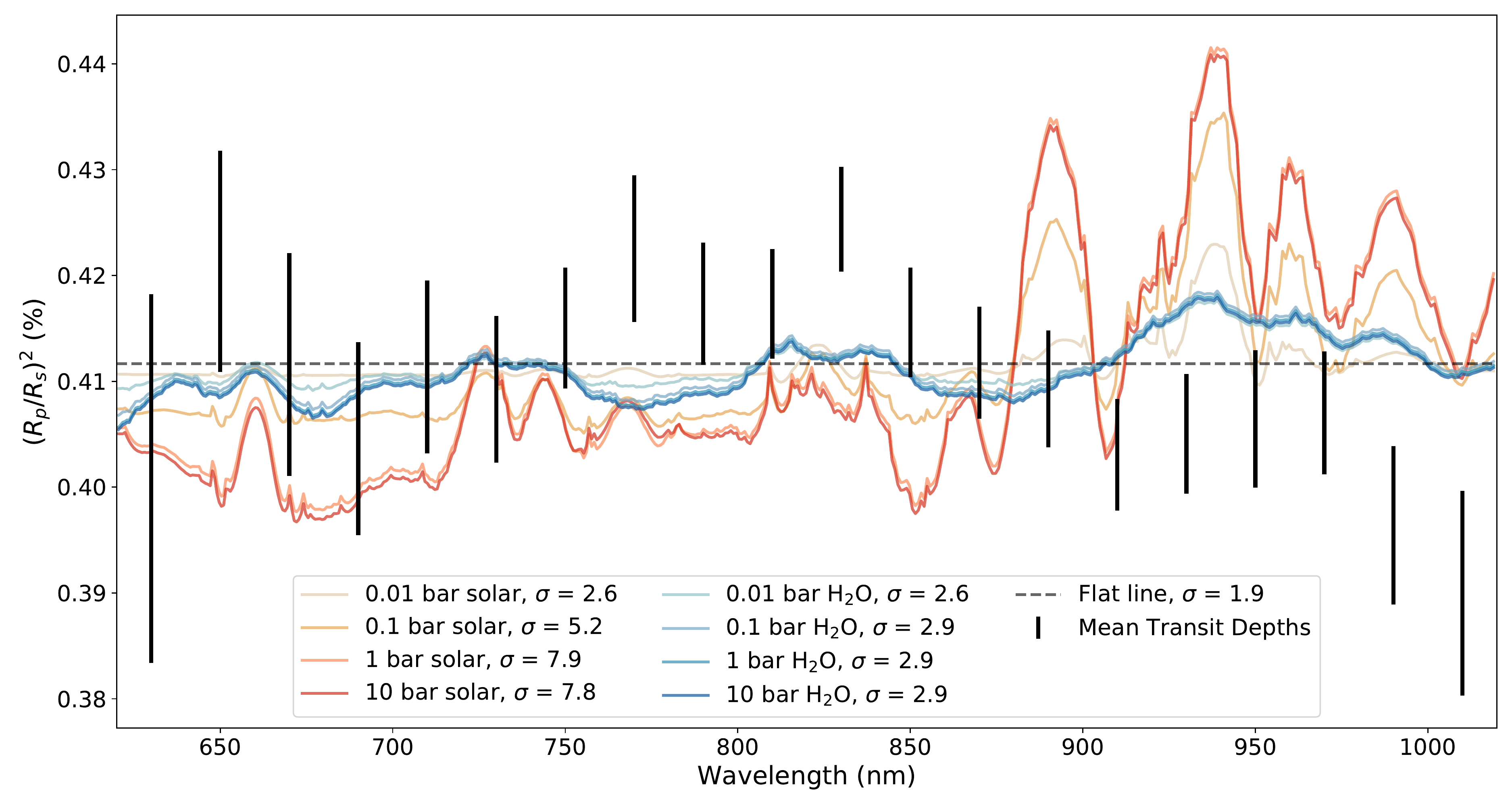}
\caption{Transmission spectrum of LHS 3844b compared to model atmospheres. The observed transmission spectrum (1-$\sigma$ black error bars) is constructed from the inverse-variance weighted mean transit depths across the 13 data sets used in this analysis. They are the same 1-$\sigma$ black error bars as in Figure~\ref{fig:transmissionhist}, and the transit depth values are provided in the grey columns of Table~\ref{tab:transitdepths}. We test two sets of clear atmospheric models: 1) solar composition, and 2) 100\% H$_2$O steam. For each model case we test surface pressures from 0.01 - 10 bar. The $\sigma$ values in the legend are the confidence to which we can rule out each model. We also compare the observed transmission spectrum to a flat line, which is more consistent with the observations.}
\label{fig:transmissionmodels}
\end{figure*}

We focus on two groups of atmospheric models of LHS 3844b: 1) a solar composition atmosphere ($\mu=2.34$), and 2) a water steam atmosphere ($\mu=18$). Within each group we test surface pressures ranging from 0.01 to 10 bars. The solar composition model is dominated by hydrogen and helium, and the main optical absorbers are water and methane. The water steam atmosphere is 100\% H$_2$O, which is such a strong absorber that in the water steam atmosphere cases it produces strong features down to 0.01 bar (Figure~\ref{fig:transmissionmodels}). We disfavor a clear, solar composition atmosphere at 0.1 bars of surface pressure and greater to 5.2 $\sigma$ confidence, while a clear H$_2$O steam atmosphere at 0.1 bars and greater is weakly disfavored at 2.9$\sigma$. We cannot rule out a flat line fit as a potential explanation of the data, meaning that our data allow for a high mean molecular weight atmosphere with low surface pressure, or no atmosphere at all. We summarize the atmospheric models we test in Table~\ref{tab:atmomodels}.

We briefly address the possibility of clouds or hazes in the atmosphere of LHS 3844b, which would truncate the transmission spectrum. \texttt{Exo-Transmit} allows the user to input a cloud-top pressure at which to place an optically-thick cloud deck in the planetary atmosphere. Because the transmission spectrum approaches a flat line as the cloud-top pressure moves to higher altitudes (lower pressures), we are able to disfavor to high confidence (5.3$\sigma$) only low-altitude clouds with cloud-top pressures of 0.1 bar and greater in the solar composition cases. We cannot rule out a high-altitude cloud deck (cloud-top pressure of 0.01 bar or less) with our observed transmission spectrum.

\begin{deluxetable}{ccccc}
\centering
\caption{Summary of atmospheric models \label{tab:atmomodels}}
\tablewidth{0pt}
\tablehead{
\multirow{2}{*}{Model} & \colhead{\textit{p}\tsub{BOA}\tsup{a}} & \multirow{2}{*}{$\mu$\tsup{b}} & \colhead{BOA adjustment\tsup{c}} & \colhead{Confidence\tsup{d}}\\
 & \colhead{(bar)} &  & \colhead{(\%)} & \colhead{($\sigma$)}
}
\startdata
\multirow{4}{*}{Solar} & 0.01 & \multirow{4}{*}{2.3} & -0.1 & 2.6 \\
& 0.1  & & -0.8 & 5.2\\
& 1.0  & & -4.3 & 7.9 \\
& 10.0 & & -9.9 & 7.8\\
\hline
\multirow{4}{*}{H$_2$O} & 0.01 & \multirow{4}{*}{18.0} & -0.2 & 2.6\\
& 0.1 & & -0.6 & 2.9\\
& 1.0 & & -1.3 & 2.9\\
& 10.0 & & -2.2 & 2.9\\
\enddata
\tablecomments{\\
\tsup{a} Pressure at the bottom of the atmosphere\\
\tsup{b} Mean molecular weight\\
\tsup{c} Adjustment made to the bottom of the atmosphere radius to scale the model atmosphere to match the observed transmission spectrum\\
\tsup{d} Confidence with which this model atmosphere is ruled out by the observed transmission spectrum}
\end{deluxetable}

\subsection{Wiggles in the LHS 3844b transmission spectrum}

Though we are able to rule out clear, low mean molecular weight atmospheres around LHS 3844b, we cannot rule out a flat line at the inverse-variance weighted mean of the transit depths ($\sigma=1.9$). We investigate the possibility that the ``wiggles'' about the mean in the observed transmission spectrum are due to inhomogeneities in the stellar photosphere, which we are effectively probing as we observe the planet transit. These inhomogeneities can arise from star spots, cooler (and darker) regions on the stellar surface, or faculae, hotter (and brighter) regions more often seen at the limb of the star \citep{Spruit1976,Foukal2004}. Both phenomena arise from magnetic activity. \citet{Rackham2018} refer to the imprint of stellar photosphere inhomogeneities on observed transmission spectra as the transit light source effect, and observed it in the optical transmission spectrum of GJ 1214b \citep{Rackham2017}, a mini-Neptune orbiting another nearby mid-M dwarf \citep{Charbonneau2009}. Star spots and faculae have temperatures different than that of the rest of the stellar photosphere, so their presence produces a chromatic effect. 

M dwarfs are known to have inhomogeneities in their photospheres, allowing for variations large enough to track their rotation periods \citep{Newton2018}. The transit light source effect in M dwarf transits can spuriously increase optical transit depths by a factor of 0$.5\% \times (R_p/R_s)^2$, with an overall slope upwards towards the blue, or spuriously decrease transit depths by a factor of $2.5\% \times (R_p/R_s)^2$ if faculae are present, with a steep downwards slope towards the blue \citep[Figures 6 \& 7 of][]{Rackham2019}. In our observed transmission spectrum we find a mean transit depth in our spectrophotometric bands of 0.4089\% and with a mean uncertainty of 0.0073\% (Table~\ref{tab:transitdepths}). A 2.5\% change in our observed transit depths would be within our 1$\sigma$ error bars. For comparison, the average transit depth of GJ 1214 in the optical is 1.3133 $\pm$ 0.0045\% \citep{Rackham2017}; the larger planet-to-star radius ratio of GJ 1214 makes changes in transit depth due to the transit light source effect larger. 

\begin{figure}
\includegraphics[width=0.49\textwidth]{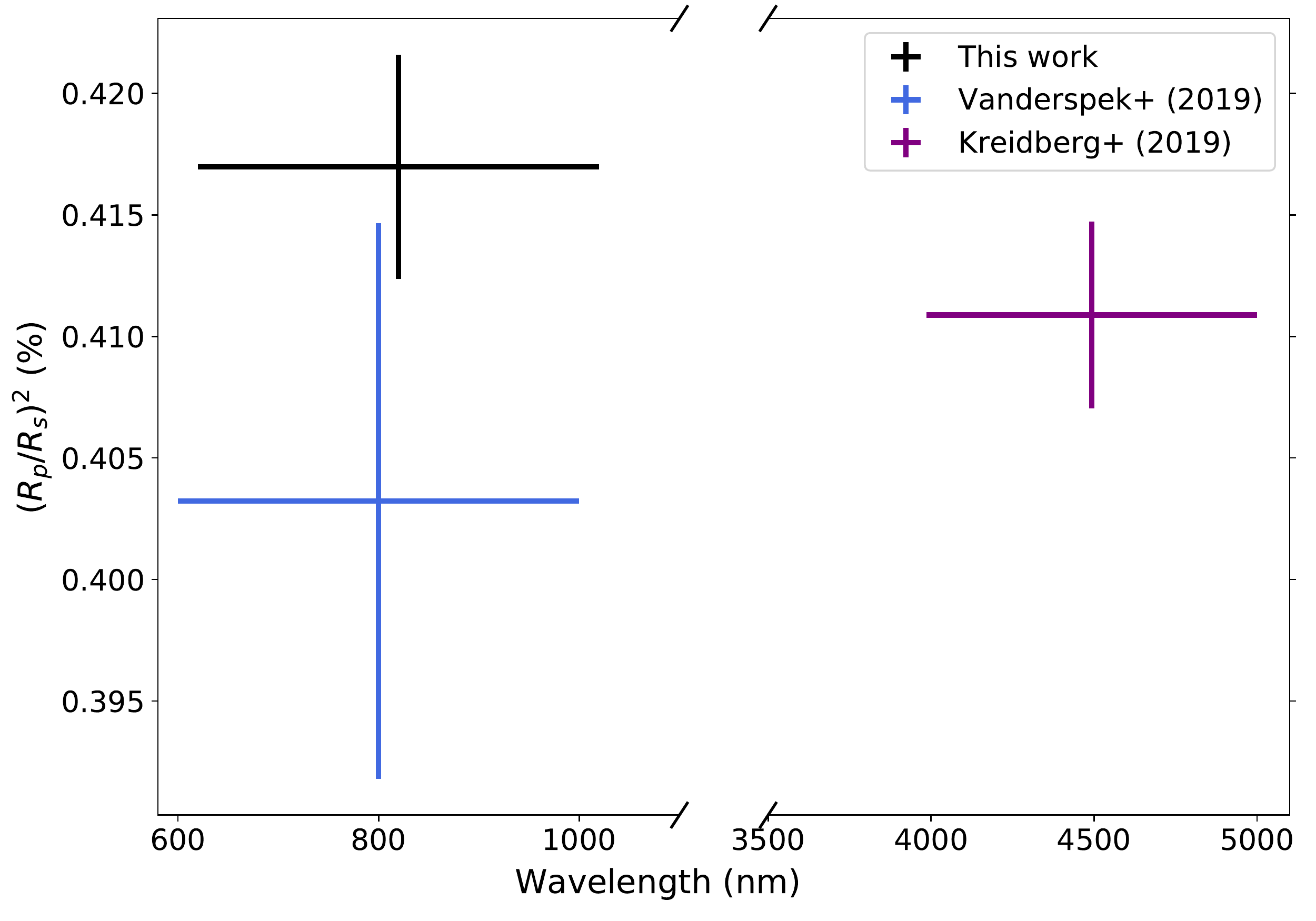}
\caption{White light curve transit depths from this work (Magellan II/LDSS3C), \citet[][TESS]{Vanderspek2019}, and \citet[][Spitzer]{Kreidberg2019}. Y-axis error bars indicate the 1-$\sigma$ uncertainties in transit depth. X-axis error bars show the extent of each bandpass.}
\label{fig:comparedepths}
\end{figure}

We focus our testing on the 10 bar solar model transmission spectrum, which has a goodness-of-fit $\chi^2=109$. Multiplying this model by stellar contamination factors according to \citet[][Table 2]{Rackham2018} does not improve the $\chi^2$. Using a fill factor for spots on the stellar photosphere of $f$\tsub{spot} = 0.9\% gives $\chi^2=118$ when compared to the observed transmission spectrum. Adjusting this spot fill factor and adding faculae such that $f$\tsub{spot} = 0.5\% and $f$\tsub{fac} = 0.46\% gives $\chi^2=109$. We also cannot find a good fit if we assume that the ``wiggles'' are produced solely by the stellar surface, and that the planetary transmission spectrum is featureless. We further note that our mean white light curve transit depth of 0.4170 $\pm$ 0.0046\% is in agreement with the transit depths found by TESS \citep[0.403 $\pm$ 0.011\%; ][]{Vanderspek2019} and Spitzer \citep[0.4109 $\pm$ 0.0038\%;][]{Kreidberg2019} for this planet (Figure~\ref{fig:comparedepths}). In particular, the Spitzer bandpass from 4-5$\mu$m should be less susceptible to the transit light source effect because the temperate differences due to photospheric inhomogeneities are minimized at longer wavelengths. 

\begin{figure}
\includegraphics[width=0.48\textwidth]{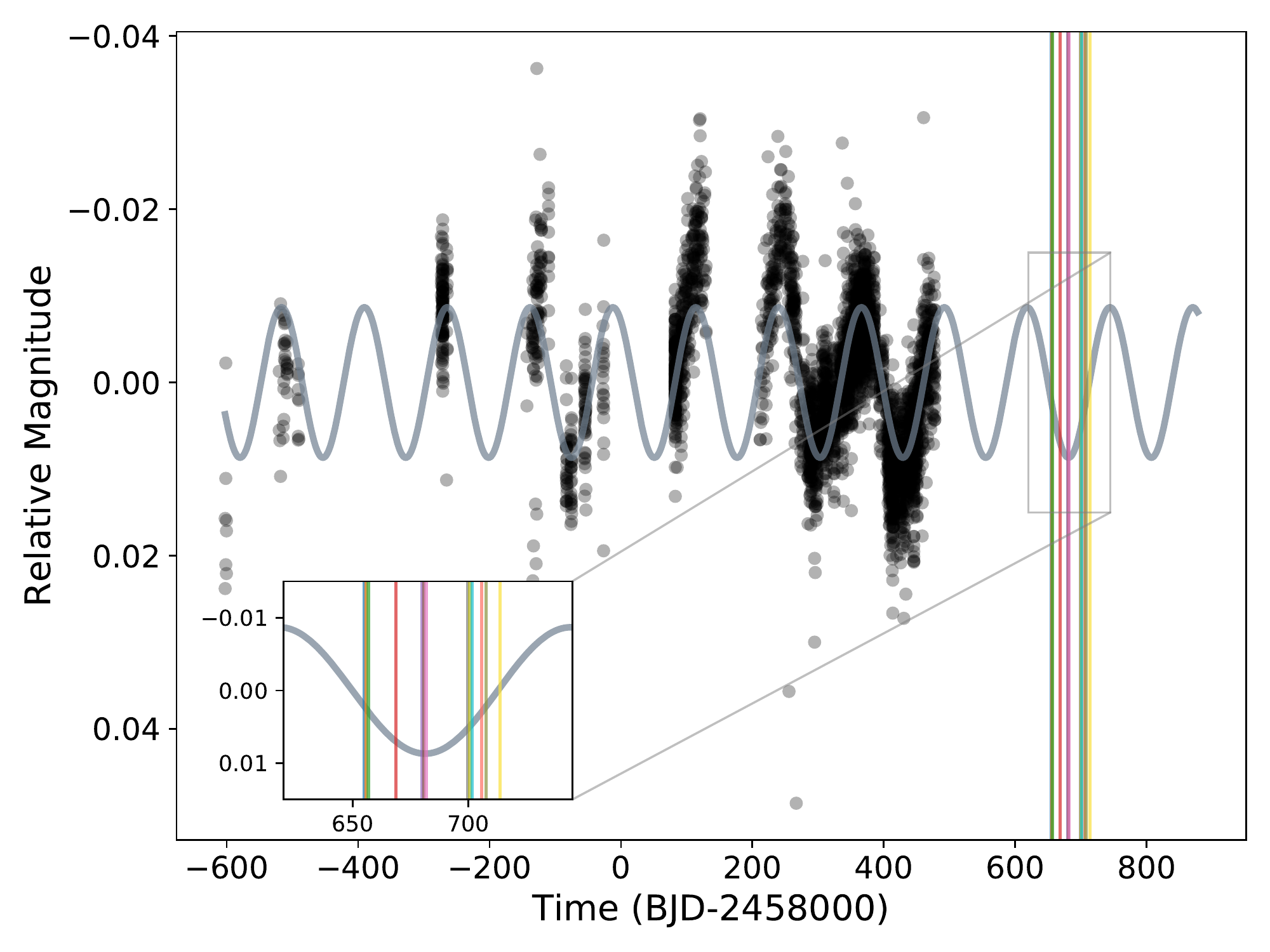}
\caption{Stellar rotation phase of LHS 3844. Black points are photometric data from MEarth South \citep{Irwin2015}. Grey line is a sinusoidal fit to the photometric data points, which was also presented in \citet{Vanderspek2019}. Vertical lines indicate the times of mid-transit for the 13 data sets analyzed in this work, with colors corresponding to those in other figures.}
\label{fig:stellarphase}
\end{figure}

Finally, we consider the stellar rotational phase over which our observations were made. The rotation period of LHS 3844 is 128 days \citep{Vanderspek2019}. The 13 transits presented here span 60 days, meaning that our data set covers half of the stellar rotation period. We did not find a correlation between white-light transit depths or slopes in the transmission spectrum with stellar rotational phase, suggesting that the transit light source effect is not apparent in our data. We propagate the rotation period of LHS 3844 forward in time to cover our observations, and find that if the rotation period is stable from 2018 to 2019, our observations sampled a valley, rather than a rapid change from peak to trough, further diminishing our chances of detecting heterogeneity in the photosphere of LHS 3844 (Figure~\ref{fig:stellarphase}).

\subsection{Comparison to previous results}

\citet{Kreidberg2019} were able to rule out atmospheres with surface pressures greater than 10 bar with their 100-hour campaign with Spitzer to observe the phase curve of LHS 3844b. By nature, those data are most sensitive to thick atmospheres which can efficiently redistribute heat from the day-side to the night-side of the planet. \citet{Kreidberg2019} argue based on theory that lighter atmospheres are not stable over the lifetime of the planet due to atmospheric erosion over time. We provide an observational constraint by addressing cases of clear, low mean molecular weight atmospheres, and disfavoring a subset of these to $>5\sigma$ confidence. 

\citet{Kreidberg2019} specifically test atmospheric compositions involving O$_2$ and CO$_2$. Models of atmospheric evolution on terrestrial planets around M dwarfs find that several bars of O$_2$ can result from hydrodynamic escape driven by high energy stellar radiation \citep{Tian2014,Luger2015,Schaefer2016}. CO$_2$ exhibits spectral features in the Spitzer Channel 2 bandpass. Unlike \citet{Kreidberg2019}, we do not address model transmission spectra comprised of O$_2$ ($\mu=32$) and CO$_2$ ($\mu=44$) because we cannot distinguish these high mean molecular weight cases from a flat line at optical wavelengths. This leaves open the possibility of a continuously replenished high mean molecular weight tenuous atmosphere around LHS 3844b.

\section{Conclusion}\label{sec:conclusion}

We observed 13 transits of the highly irradiated terrestrial exoplanet LHS 3844b in the fall of 2019 with the Magellan II (Clay) telescope and the LDSS3C multi-object spectrograph at the Las Campanas Observatory in Chile. From these 13 transits we construct both white light curves and spectroscopic light curves. When combining all 13 data sets we achieve an RMS precision of 112 ppm in 3-minute time bins of the white light curve, and an RMS of 65 ppm if we bin down to 10-minutes. We derive a combined value of $(R_p/R_s)^2$ = 0.4170 $\pm$ 0.0046\%. 

We chop the light curves into 20 spectrophotometric bands of 20 nm each. We take the inverse-variance weighted mean of the 13 transit depths in each band to construct our combined transmission spectrum. We achieve an average transit depth precision on $(R_p/R_s)^2$ of 0.0073\%, and a median of 1.64 $\times$ the expected noise in the spectroscopic light curves. We compare the final transmission spectrum to models of LHS 3844b's atmosphere. We exclude clear low mean molecular weight solar composition atmospheres with surface pressures of 0.1 bar and greater to $5.2\sigma$ confidence, and clear, 100\% H$_2$O water vapor atmosphere with surface pressures of 0.1 bar and greater to $2.9\sigma$ confidence. In the case of solar composition atmospheres, we rule out clouds with cloud-top pressures of 0.1 bar and greater to 5.3$\sigma$ confidence but we cannot address clouds at lower pressures (higher altitudes).

Our results are in good agreement with theoretical models and observational evidence demonstrating that terrestrial worlds do not retain low mean molecular weight atmospheres \citep{deWit2016,deWit2018,Diamond-Lowe2018}. The question remains if terrestrial exoplanets orbiting M dwarfs can retain thick, high mean molecular weight atmospheres, as the Solar System terrestrial planets do. In the case of the highly irradiated planet LHS 3844b, this work and the previous study by \citet{Kreidberg2019} indicate that the answer is likely no. But cooler worlds in the growing sample of nearby terrestrial exoplanets orbiting low-mass stars may prove differently. These cooler terrestrial exoplanets are not spectroscopically accessible to us today, but the next generation of space-based observatories beginning with the James Webb Space Telescope, and ground-based telescopes like the Giant Magellan Telescope, the Thirty Meter Telescope, and the European Extremely Large Telescope, will be able to characterize the atmospheres, or lack thereof, around these worlds.

\acknowledgments
This paper includes data gathered with the 6.5m Magellan II Telescope (Clay) located at Las Campanas Observatory, Chile. We thank the contributors to the LDSS3C instrument and documentation, the telescope operators and staff at Las Campanas Observatory, and the writers and contributors of the open-source software used in this work. We make use of the Digitized Sky Surveys in Figure~\ref{fig:onsky} of this paper, as well as during observations. The Digitized Sky Surveys were produced at the Space Telescope Science Institute under U.S. Government grant NAG W-2166. The images of these surveys are based on photographic data obtained using the Oschin Schmidt Telescope on Palomar Mountain and the UK Schmidt Telescope. The plates were processed into the present compressed digital form with the permission of these institutions. We thank Jonathan Irwin for providing MEarth photometric data of LHS 3844 and the model for stellar rotation. H.D.-L. recognizes support from the National Science Foundation Graduate Research Fellowship Program (grant number DGE1144152). M.M. acknowledges support from the Swiss National Science Foundation under the Early Postdoc Mobility grant P2BEP2\_181705. This publication was made possible through the support of a grant from the John Templeton Foundation. The opinions expressed here are those of the authors and do not necessarily reflect the views of the John Templeton Foundation. This material is based upon work supported by the National Aeronautics and Space Administration under Grant No. 80NSSC18K0476 issued through the XRP Program.

\facilities{Magellan II (Clay), LDSS3C}

\software{\texttt{astropy} \citep{AstropyCollaboration2013,AstropyCollaboration2018}, \texttt{batman} \citep{Kreidberg2015}, \texttt{decorrasaurus} \citep{Diamond-Lowe2020}, \texttt{dynesty} \citep{Speagle2020}, \texttt{Exo-Transmit} \citep{Kempton2017}, \texttt{george} \citep{Foreman-Mackey2015}, \texttt{LDTk} \citep{Parviainen2015}, \texttt{mosasaurus} (\href{http://www.github.com/zkbt/mosasaurus}{github.com/zkbt/mosasaurus}), \texttt{SAOImageDS9} \citep{Joye2003}}

\medskip
\bibliography{MasterBibliography.bib}

\end{document}